\documentstyle[prd,aps,epsf,floats]{revtex}
\draft
\newcommand{\lsim}{\mathrel{\mathop{\kern 0pt \rlap
  {\raise.2ex\hbox{$<$}}}
  \lower.9ex\hbox{\kern-.190em $\sim$}}}

\newcommand{\be}{\begin{equation}}
\newcommand{\ee}{\end{equation}}
\newcommand{\ba}{\begin{eqnarray}}
\newcommand{\ea}{\end{eqnarray}}
\newcommand{\nn}{\nonumber}
\newcommand{\ep}{\epsilon}
\newcommand{\bbn}{big-bang nucleosynthesis}

\newcommand{\plb}{Phys. Lett. B$~$}

\newcommand{\npb}{Nucl. Phys. B}
\begin{document}

\twocolumn[\hsize\textwidth\columnwidth\hsize\csname
@twocolumnfalse\endcsname

\title{
\hbox to\hsize{\large Submitted to Phys. Rev. D. \hfil E-print: Astro-ph/9800000}
\vskip 1.0cm
Weak reaction freeze-out constraints on primordial magnetic fields
}

\author{
In-Saeng Suh 
\footnote{E-mail: isuh@cygnus.phys.nd.edu}
and G. J. Mathews
\footnote{E-mail: gmathews@bootes.phys.nd.edu}
}


\address{
Department of Physics,
University of Notre Dame,
Notre Dame, Indiana 46556, USA 
}

\date{\today}
\maketitle   
 
\begin{abstract}
We explore constraints on the strength of the primordial magnetic field based upon
the weak reaction freeze-out in the early universe.
We find that limits on the strength of the magnetic field found in other works are
recovered simply by examining the temperature at which the rate of weak reactions 
drops below the rate of universal expansion ($\Gamma_{w} \le$ H).
The temperature for which the $n/p$ ratio at freeze-out leads to acceptable helium
production implies limits on the magnetic field.
This simplifies the application of magnetic fields to other cosmological
variants of the standard big-bang.
As an illustration we also consider effects of neutrino degeneracy on the 
allowed limits to the primordial magnetic field.
\end{abstract}

\vskip 1.2pc]

\narrowtext

\section{Introduction}

It is by now widely recognized that cosmological magnetic fields could be a 
significant factor on almost every scale relevant to the structure and evolution 
of the universe.
Recent measurements \cite{beck,kim} of intergalactic magnetic fields have provided
evidence that
magnetic fields on the order of a few $10^{-6}$G are ubiquitous, e.g. in numerous 
galaxies, galactic halos, and clusters of galaxies. 
Similar large magnetic field strengths have also been found in protogalactic clouds
\cite{lanzetta}.
However, the origin of these fields and the existence of galactic and inter-cluster
magnetic fields are still an issue of debate \cite{olinto,enqvist,barrow}. 
For a long time, it has been supposed that a turbulent
dynamo mechanism \cite{zeldovich} may be at work in which the magnetic fields might have
arisen from the
exponential amplification of small seed fields by hydrodynamic turbulence.
However, recent detailed models \cite{goldman} show that this dynamo model is inadequate.
The fields that can be generated in this process are an order of magnitude
weaker than what is actually observed. 
Goldschmidt and Rephaeli \cite{goldshmidt} also showed that Faraday rotation
measurements could not be explained by a turbulent dynamo model.
Furthermore, the existence \cite{lanzetta} of damped Ly-$\alpha$ lines in QSO's
indicates that primordial magnetic fields existed at the early times.

Faraday rotation measure (RM) is defined by the rotation angle of polarized light in
a magnetic field. It depends upon the strength and spatial extent of the magnetic
field, the density of the associated
plasma, and on the wavelength of the observed radiation \cite{zeldovich}. Therefore,
the observation of Faraday rotation in radio sources both inside and behind clusters
gives information on the strength of the inter-galactic magnetic field \cite{kim}.
For example, the detected magnetic field in clusters of galaxies has a typical magnitude
of a few $10^{-6}$G and a coherence length scale of $10 \sim 100$ kpc \cite{taylor}. 
Another important estimate of the cosmological magnetic field has been established by 
Kosowsky and Loeb \cite{kosowsky}. 
They argue that the existence of primordial magnetic fields 
at the last scattering surface may induce a measurable Faraday rotation in the
polarization of the cosmic microwave background radiation. According to their results,
it should be possible to detect the presence of a magnetic field at photon decoupling
which would corresponds to a present day magnetic field as small as $B = 10^{-9}$G.

General effects of magnetic fields in astrophysical and cosmological processes 
have been investigated by several authors
\cite{langer,green,oconnell,cheng,grasso,kernan}.
In particular, the effect of a primordial magnetic field on \bbn~was first studied by
Greenstein \cite{green} and Matese and O'Connell \cite{oconnell}.
They argued that if primordial magnetic fields of sufficient strength existed in the
early universe, then these could have had direct influences on both the expansion rate
of the universe and the nuclear reaction rates \cite{green,oconnell,cheng,grasso,kernan}.
These influences also could affect weak reaction freeze-out and hence the
abundances of the light elements produced during the primordial \bbn~epoch.

The weak interaction rates themselves can also be affected by magnetic fields.
This can influence significantly the rate of production of $^4He$ and other light
elements.
Recently, Cheng, et al. \cite{cheng} have considered the effects of a primordial
magnetic field on \bbn. In that work they used weak interaction rates 
modified by the presence of magnetic fields.
They obtained constraints on the maximum strength of the primordial
magnetic field in the framework of standard \bbn.
According to their results, the allowed primordial magnetic field intensity
at the end of nucleosynthesis ($T \approx 10^8$ K) is about
$B \lsim 2 \times 10^9 - 10^{11}$ G on scales greater than $10^{4}$ cm
but smaller than the event horizon at the \bbn~ epoch.
Subsequently, Grasso and Rubinstein {\cite{grasso}} obtained an upper limit
of $B \leq 10^{12}$ G if the coherence length $L_0$ of the magnetic field 
at the end of primordial nucleosynthesis is in the range of $10 << L_0 << 10^{11}$ cm. 
In another work, Kernan, et al. \cite{kernan} obtained an upper limit of $B \le 1
\times 10^{11}$G at $T=10^9$ K. 

The main purpose of the present letter is to point out that the essential
results of those papers can be understood from the condition of weak reaction
freeze-out alone and not on the later nucleosynthesis epoch.
This is important for two reason: one is that the magnetic field strength
is being constrained at an earlier time (1 sec rather than 100 sec);
the other is that the constraint can be easily obtained without 
the necessity to run a full nucleosynthesis computation.
This will allow easy parameter studies of effects of magnetic fields
on a wide variety of cosmological models. As an illustration we consider a 
neutrino degenerate model here.

During early epochs, particle species and fundamental interactions
depart from equilibrium as the temperature of the
universe decreases \cite{padman,kolb}.
One of the typical departures is that of weak interaction decoupling which occurs
when the temperature of the universe was about 1 \,MeV.
Weak-interaction freeze-out has important direct effects on the evolution of the
universe. The synthesis of light elements depends sensitively upon the 
neutron to proton ratio $n/p$ at the start of the nucleon reaction 
epoch ($T \lsim 10^9$ K).
However, the $n/p$ ratio is determined by the competition 
between the weak interaction rates and the expansion rate of the universe.
A higher equilibrium $n/p$ ratio at weak-interaction freeze-out will lead to 
a higher $^4 He$ abundance during \bbn. 

In this work we constrain the strength of magnetic fields at the 
weak reaction freeze-out epoch, without considering how such fields could have formed.
Our purpose is, therefore, to estimate the maximum allowed strength of the primordial
magnetic field which is still consistent with  weak-interaction freeze-out 
at around 1 MeV. In order to determine the maximum allowed strength of a magnetic field 
at this epoch, 
we here estimate the change in weak interaction rates in the presence of a magnetic 
field as well as the change in the Hubble expansion rate from the presence of 
magnetic energy density. 
In addition, we consider the possible effects of neutrino degeneracy on the
primordial magnetic field.
Although we calculate the main effects on weak reactions, we have neglected
the much smaller effects from some higher order interactions, radiative corrections, and
other medium effects
on the weak interactions. The electron mass  
will also be changed in strong magnetic fields, i. e. $m_{e}(B) =
m_{e}(B=0) [1 + {\cal O}(\alpha_e)
+ \cdots]$, with $\alpha_e \simeq 1/137$ being the electron fine structure constant
\cite{gusynin}, but for the magnetic fields of interest, 
other terms higher than ${\cal O}(\alpha_e)$ are negligible.
We employ a system of units in which $\hbar = k_B = c = 1$, except when specific units
must be attached to a result.
 
\section{Expansion rate and weak reaction rates in the presence of a magnetic field}

The properties of an electron in an external magnetic field have been extensively 
studied in a number of papers \cite{gusynin,schwinger,johnson,landau}. 
In brief, the energy states of the electron in a
magnetic field are quantized and the properties of an electron are modified
accordingly. In order to investigate the properties of an electron in a magnetic field,
we must first solve the Dirac equation in an external static and homogeneous
magnetic field. We make the convenient choice of gauge for the vector potential in which 
a uniform magnetic field $B$ lies along the $z$-axis. 
We then obtain the electron wavefunctions in a magnetic field as was calculated by
Johnson and Lippmann in detail \cite{johnson}.
The dispersion relation for an electron propagating through
a magnetic field is \cite{cheng}  
\be
E = [p_z^2 + m_e^2 + 2eBn_s]^{1\over 2} + m_e \kappa ,
\ee  
where $n_s= n+{1\over 2}-s_z , \ (n_s=0, 1, \, ...$), 
$n$ is the  principal quantum number of the Landau level, 
$s_z =\pm 1/2$ are the electron spins, $e$ is the electron charge, 
$p_z$ is the electron momentum along the $z$-axis,
$m_e$ is the rest mass of the electron, and $\kappa$ is the anomalous
magnetic moment for an  electron in the ground state ($n=0, \, s_z=1/2$).  
For relatively weak fields (i. e.,  $B \lsim 7.575\times 10^{16}$ G),
$\kappa = -\frac{\alpha_e}{4\pi} \frac{eB}{m_e^2}$,  
while for stronger fields, $\kappa = \frac{\alpha_e}{2\pi} (\ln \frac{2eB}{m_e^2})^2$ 
\cite{schwinger}.
However, the analyses available in the literature \cite{gusynin} indicate that these
higher order corrections have a very small effect at the magnetic field
strengths and temperatures of interest. Thus, we can ignore the anomalous magnetic 
moment term.
Strictly speaking, all states of the neutron and proton are affected in the presence of 
a magnetic field. However, the effect is also smaller by a factor $(m_e / M_p)^2$.
Hence, the modification of the proton and neutron states by
a magnetic field can also be neglected.

The main modification of the electrons in a magnetic field comes from the available
density of states for the electrons \cite{landau}. 
The electron state density in the absence of a magnetic field is 
\be
2 \int \frac{d^{3} p}{(2 \pi)^3}.
\ee
In a magnetic field this is replaced with
\be
\sum_{n_s =0}^{\infty}[2-\delta_{n_s 0}] \int {eB\over(2\pi)^2} dp_z.
\ee
This modification will affect the thermodynamic variables for the 
electrons as well as the fundamental interaction rates. 

Now let us consider the effects of a magnetic field on the expansion rate. 
If the primordial magnetic field is spread over sufficiently small distances
compared with the event horizon, the geometry of the universe
is not affected \cite{green,turner}. Thus, a Robertson-Walker metric is still 
appropriate and 
the expansion rate of the universe can be simply described 
by the usual Friedmann equation,
\be
H = \frac{1}{R} \frac{d R}{d t} = \sqrt{\frac{8 \pi}{3} G \rho},
\ee
where $G$ is the gravitational constant and $\rho$ is the total mass-energy density
\be
\rho = \rho_{\gamma} + \rho_{e} + \rho_{\nu} + \rho_{b} + \rho_{B}.
\ee
The subscripts $\gamma, e, \nu, b$, and $B$ denote the mass energy due to photons,
electrons, neutrinos, baryons, and the magnetic field, respectively.
Thus, the presence of a magnetic field changes the expansion rate in two ways. One is
by the associated magnetic energy density 
\be
\rho_{B} = \frac{B_{c}^{2}}{8 \pi} \gamma^2 ,
\ee
where $\gamma = B/B_c$ and $B_{c} = m^{2} /e =4.414 \times 10^{13}$ G is the critical 
magnetic field at which quantized cyclotron states begin to exist.
The other is the modification of the electron mass-energy density.
Since the phase space of electrons in a magnetic field is modified,
the electron energy density is given by 
\ba
\rho_e(B) &=& 2 \frac{m_e^4}{(2 \pi)^2} \gamma   
\sum_{n_s=0}^{\infty}[2-\delta_{n_s0}]
\int_{\sqrt{1+2\gamma n_s}}^\infty d\ep 
\frac{\ep^2}{1 + e^{\ep a + \phi_e}} \nn\\
&~&\times \frac{1}{\sqrt{\ep^2-(1+2\gamma n_s)}} ,
\ea
where $\ep = E /  m_e, \;  a = m_e / T_e$, $\phi_e = \mu_e / T_e$, and
$T_e$ is the temperature of the electrons. 
Since the electron chemical potential is small in the early universe, i.e., 
$\phi_e \lsim 10^{-9}$ \cite{kolb}, we can ignore it in the calculations.


The weak reaction rates in a magnetic field have been derived by several
authors \cite{oconnell,cheng,grasso,kernan}.
In thermal equilibrium, the inter-conversion between neutrons and protons is
possible through the weak reaction processes:
\ba
n + \nu &\longleftrightarrow& p + e^{-} \\
n + e^{+} &\longleftrightarrow& p + \bar{\nu} \\
n & \longleftrightarrow& p + e^{-} + \bar{\nu}
\ea
These weak processes set the $n/p$ ratio in various astrophysical
processes such as \bbn \cite{padman,kolb}, neutron star cooling \cite{shapiro}, etc.
The reaction rate for each process can be calculated using the well-known V-A theory.  
The total weak reaction rates for the conversion of neutrons into protons 
in an external magnetic field is written as  
\ba
\Gamma_{n \to p} (B) &=& \frac{\gamma}{\tau} \sum_{n_s=0}^{\infty}
[2-\delta_{n_s0}] \int_{\sqrt {1+2\gamma n_s}}^\infty d\epsilon
\frac{\ep}{1+e^{\ep a}} \nn \\
&~& \times \frac{1}{\sqrt{\ep^2 - (1+2 \gamma n_s)}}  
\bigg[ \frac{(\ep+q)^2 e^{b (\ep + q) + \xi_e}}{1+e^{b (\ep + q) + \xi_e}} \nn\\
&~& + \frac{(\ep - q)^2 e^{\ep a}}{1+e^{b (\ep - q) - \xi_e}} \bigg], 
\ea
where $1/\tau \equiv g_{V}^{2} (1 + 3 \alpha^2) m_e^5 / 4 \pi^3
\simeq 3.26 \times 10^{-4}$ sec$^{-1}$, 
$g_V = 1.4146 \times 10^{-49}$ erg cm$^{3}$, and
$\alpha = g_{A}/g_{V} \simeq -1.262$.
The parameters used in Eq. (11) are defined as
$q=Q / m_e$, $b = m_e / T_\nu$, and $\xi_e = \mu_{\nu_e} / T_{\nu}$. 
The quantity $Q = M_n - M_p \simeq \; 1.293$ MeV is the neutron-proton mass difference. 
$\xi_e$ is the electron neutrino degeneracy parameter which remains constant in the
expanding universe,
$\mu_{\nu_e}$ is the chemical potential of electron neutrino, and  
$T_{\nu}$ is the neutrino temperature. 
Here, we neglect the polarization of the neutron source \cite{oconnell}. 

The inverse total reaction rate for the conversion of protons to neutrons can
be obtained from detailed balance.
\be
\Gamma_{p \to n} (B) = e^{-q a -\xi_e} \Gamma_{n \to p} (B). 
\ee
In the limit of vanishing magnetic field ($\gamma \rightarrow 0$), 
$\Gamma_{n \to p}(B)$ reduces to $\Gamma_{n \to p}(B=0)$ \cite{kernan}. 
The weak interaction rates, Eqs. (11) and (12), increase 
as the field strength $\gamma$ increases for 
fixed temperature (Ref. \cite{grasso} and see Fig. 1). 

From the above we see that magnetic fields affect nucleosynthesis in two 
competing ways, through the expansion rate of the universe H 
and the weak reaction rates $\Gamma_w$. 
The magnetic energy density $\rho_B$ in the total energy density accelerates 
the expansion of the universe and
therefore increases the primordial $^4$He abundance. 
At the same time, however, the weak reaction rates increase in a magnetic field.
This extends weak-reaction equilibrium to lower temperature which reduces the $^4$He
abundance.
  
\section{weak-reaction freeze-out in the absence of the neutrino degeneracy}

As long as the reaction rates exceed the expansion rate, chemical
equilibrium is obtained. The equilibrium ratio $n/p$ in the absence of the neutrino
degeneracy is then given by \cite{padman,kolb}
\be
\frac{n}{p} = exp(-Q/T).
\ee
This ratio is maintained only as long as the reaction rate per baryon $\Gamma$ 
remains greater than the cosmic expansion rate $H$. 
Then, at some temperature $T_D$, the weak reactions decouple.
The neutron to proton ratio $n/p$
subsequently remains nearly frozen at its equilibrium value $exp(-Q/T_D)$.
The $n/p$ ratio does, however, slowly decrease due to occasional weak reactions.
Eventually the ratio is dominantly affected by the beta decay of free neutrons which
continue to decrease this ratio until all of the 
neutrons are bound into nuclei.
The freeze-out temperature $T_D$ of the weak interactions is essentially
determined by the condition $\Gamma(T_D) = H(T_D)$ for which one obtains
$T_D \approx 0.75$ MeV in the absence of magnetic fields \cite{padman,kolb}.

In big-bang nucleosynthesis, the abundance of $^4He$ can be easily estimated by 
assuming that almost all neutrons are incorporated into $^4He$. 
The resulting mass fraction of $^4He$ will be approximately
\be
Y_p \simeq \frac{2 (n/p)_{BBN}}{1 + (n/p)_{BBN}},
\ee
where $(n/p)_{BBN}$ is approximately determined by the equilibrium value of
the neutron to proton ratio $n/p$ and the neutron decay factor prior to deuterium
formation.
However, in the presence of strong magnetic fields (i.e., $\gamma \sim 1$), 
the neutron decays more rapidly (about 15 $\%$ faster) than in the field-free case 
\cite{fassio}. The neutron decay factor will thus decrease and  
the allowed freeze-out temperature $T_D$ from the primordial $^4$He constraint
will increase.
Using this we can approximately estimate the allowed values of $(n/p)_{BBN}$ 
in the presence of primordial magnetic fields during \bbn. 
For example, adopting $Y_p \simeq 0.25$, we find $T_D \simeq 0.77$ MeV.

Now if the weak reaction freeze-out temperature $T_D$ is fixed by the value of 
primordial $Y_p$, we can obtain a limit on the primordial magnetic field strength 
at that epoch. 
In order to constrain the strength of the magnetic field we again use the condition
\be
\Gamma (B_D) = H (B_D),
\ee
for a given weak reaction freeze-out temperature $T_D$.
In Eq. (15), $B_D$ denotes the magnetic field strength when the weak reactions 
decouple. Therefore $\gamma_D = B_D / B_c$.

Figure 1 shows the weak interaction rates from neutrons converting to protons 
$\Gamma_{n \to p}$ and the expansion rates $H$  as a function of   
$\gamma = B/B_c$ for the given temperatures $T = 0.6, 0.8, 1.0$ and 1.2 MeV
respectively and $\xi_e = 0$. 
For a given temperature, the expansion rates are seriously affected
by magnetic fields for $\gamma \ge 1$. On the other hand, large magnetic fields
($\gamma > 10$) significantly affect the weak reaction rates.   
For $T_D \ge 0.75$ MeV, we can find the intersection points of $H$ and $\Gamma$. 
These are the magnetic field strengths $\gamma_D$ at which the weak reactions
freeze-out. As a result the freeze-out temperature increases as $\gamma$ increases.  


\section{weak-reaction freeze-out in the presence of the neutrino degeneracy}

As an illustration of how the above analysis is easily applied to other cosmological
paradigms, we now consider the effect of a possible chemical potential for the 
electron neutrino, $\mu_{\nu_e}$.
In this case the distribution functions of the neutrinos are different from   
those of the anti-neutrinos.
Here, we choose to ignore the muon and tau ($m_{\mu}$ and $m_{\tau} >>$ 1 MeV)
chemical potential because at the nucleosynthesis epoch they have already disappeared 
through annihilation and decay.

By this time there is no experimental or direct theoretical constraints for the
magnitude or sign of neutrino asymmetries. However, there exist indirect 
constraints on the neutrino degeneracies obtained from primordial nucleosynthesis
\cite{kang,mathews}.
Including neutrino degeneracy has two effects: (1) the energy density
of neutrinos increases, the expansion rate of the universe thus increases;
(2) weak reaction rates are modified by the change in the electron-neutrino
distribution functions. Therefore the non-vanishing electron-neutrino degeneracy
can directly effect the equilibrium $n/p$ ratio at weak reaction freeze-out.

With such a nonzero chemical potential of the electron neutrinos,
the energy density of neutrinos is given by \cite{kang}
\be 
\frac{\rho_{\nu}}{\rho_{\gamma}} = \frac{21}{8} \left(\frac{T_{\nu}}{T}\right)^{4} 
\left[1 + \frac{10}{7} \left(\frac{\xi_e}{\pi}\right)^{2} 
+ \frac{5}{7} \left(\frac{\xi_e}{\pi}\right)^{4} \right].
\ee
The equilibrium $n/p$ ratio is also related to the $\xi_e$ by
\be
\frac{n}{p} = exp[-Q/T - \xi_e] .
\ee
Eventually, increasing $\xi_e$ leads to a smaller value of $n/p$ when the weak
reaction rates freeze out and hence a smaller production of primordial helium.

There have been various estimates for the constraint on the neutrino degeneracy 
from big-bang nucleosynthesis \cite{kang,sato,kimj}. With the constraints imposed
by large scale structure formation, Kang and Steigman obtained the limits on the
electron neutrino degeneracy of $-0.06 \lsim \xi_e \lsim 0.15$ \cite{kang}.   
Recently Ref. \cite{sato} has shown that big-bang nucleosynthesis calculations agree
with the primordial abundances of light elements inferred from the observational
data if the electron neutrino has a chemical potential due to lepton asymmetry. 
They obtain the possible constraints of neutrino degeneracy,
$0.003 \le \xi_e \le 0.083$ for $3.1 \le \eta_{10} \le 5.5$, 
where $\eta_{10}= \eta \times 10^{10}$, $\eta$ is the baryon-to-photon ratio $n_b /
n_{\gamma}$. 
Similarly, Kim et al. \cite{kimj} obtained $|\xi_e| \simeq 1$ with recent updated 
constraints on primordial light elements.  

In this work, therefore, we will adopt $\xi_e =0.15$ to see the maximum effect of 
neutrino degeneracy on the constraint of primordial magnetic field.
We calculate Eq. (15) in both cases of $\xi_e =0$ and $\xi_e = 0.15$.
Figure 2 shows the freeze-out temperature $T_D$ versus $\gamma_D$ 
at the weak reactions freeze-out.
We can see that $\gamma_D$ goes to zero as $T_D$ approaches to 0.75 MeV
and there is no $\gamma_D$ for $T_D < 0.75$ MeV.
In the case of $\xi_e = 0$,, we can estimate the freeze-out temperature 
$T_D \lsim 0.76$ MeV which satisfies the primordial $^4$He constraint, 
$Y_p \lsim 0.245$ \cite{copi}. 
Therefore, we obtain $\gamma_D \lsim 0.14$ for $T_D \lsim0.76$ MeV. 
In the case of $\xi_e = 0.15$, the neutrino degeneracy raises the freeze-out 
temperature by $\Delta T_D \sim 0.025$ MeV at $\gamma \lsim 1$.  
However, the $n/p$ ratio for the positive $\xi_e$ is suppressed,
hence $Y_p$ is reduced. 
Therefore we can allow a higher weak freeze-out temperature
$T_D \lsim 0.82$ MeV (corresponding to $\gamma_D \lsim 0.8$) and still
satisfy the constraint, $Y_p \lsim 0.245$.
For the more stringent constraint $\xi_e \le 0.083$, we have $T_D \lsim 0.78$ MeV. 
Thus, we obtain $\gamma_D \lsim 0.2$ for $T_D \lsim 0.78$ MeV and
it is possible to have $\gamma_D \lsim 0.8$ for $T_D \lsim 0.82$ MeV. 

\section{Conclusions}

In this work, we have shown that weak-interaction freeze-out in the early universe 
is sufficient to provide a constraint on the strength of the primordial magnetic field.
If magnetic fields existed in the early universe,
in particular, during  the weak reaction freeze-out, 
they could have influenced  both the expansion rate of the universe
and the weak reaction rates. Therefore the weak reaction freeze-out temperature
would be changed with respect to the field-free case.
Subsequently, these influences affect the abundances of light elements which
are produced in this magnetized environment.
A chemical potential in electron neutrinos can change the equilibrium 
$n/p$ ratio [Eq. (17)]
as well as increase the expansion rate of the universe. We also consider the
possible effects of neutrino degeneracy on the primordial magnetic field limits.

In order to determine the maximum allowed strength of any primordial magnetic field, 
we use the simple condition, $\Gamma (B_D) = H (B_D)$ for the  given freeze-out
temperatures $T_D$.
Since the observational requirement for the primordial $^4He$ abundance is 
that $Y_p \lsim 0.245$ \cite{copi},
the weak freeze-out temperature $T_D$ should be less than about 0.76 MeV for
$\xi_e = 0$ or 0.82 MeV for $\xi_e = 0.15$. 
We therefore find an upper limits from our results of
\be
B \lsim 6.2 \times 10^{12}\,G 
\ee
for the weak reaction freeze-out temperature $T_D = 0.76$ MeV ($\xi_e = 0$)
and 
\be
B \lsim 3.5 \times 10^{13}\,G         
\ee
for the weak reaction freeze-out temperature $T_D = 0.82$ MeV ($\xi_e = 0.15$).

Since the universe has been a good conductor through most of its evolution,
the cosmic magnetic field will conserve magnetic flux as it evolves \cite{turner}. 
We therefore can obtain a simple relation
\be
B \propto R^{-2} \propto T^{2}.    
\ee
If we assume that the cosmic magnetic field has continued to rescale 
according to Eq. (17), 
our results imply that the present cosmic magnetic field is less than 
$5.8 \times 10^{-7}$G ($\xi_e = 0$) and $2.8 \times 10^{-6}$G ($\xi_e = 0.15$). 
The latter is larger than the limit determined in Ref. \cite{grasso} ($3 \times 10^{-7}$G)
by a factor 10 but the former is nearly similar. 

In this work we have shown that a constraint on the strength of primordial 
magnetic field can be inferred by weak-interaction freeze-out in the early 
universe without full calculations of \bbn.
Similarly, another important decoupling phenomenon in the early universe is the neutrino
decoupling which occurred about $T\simeq$ 1.4 MeV.
Neutrino decoupling is also very significant to the evolution of the early universe.
Clearly, this neutrino decoupling will provide a constraint on the strength of 
primordial magnetic field before \bbn. This will be 
the subject of a forthcoming paper \cite{isuh}. 

\vspace{0.5cm}
 
{\bf Acknowledgments.}
\noindent
I.S.S. acknowledges the Korea Research Foundation (KRF) for financial support.
This work supported in part by DOE Nuclear Theory Grant DE-FG02-95ER40934.
 


 
\vskip 2cm
\newpage
\begin{center}
\begin{large}
FIGURE CAPTIONS
\end{large}
\vspace{5mm}\
\end{center}
 
\begin{itemize}
\item [{\rm FIG. 1}]
The total weak interaction rate for converting neutrons to protons $\Gamma_{n \to p}$
and the expansion rates $H$ as a function of $\gamma = B/B_c$ 
are plotted for the values of freeze-out temperatures 
$T_D = 0.6, 0.8, 1.0,$, and 1.2 MeV and $\xi_e = 0$ as noted.

\item [{\rm FIG. 2}]
Freeze-out temperature $T_D$ versus the magnetic field strength
$\gamma_D$ at the weak interaction freeze-out. 
The solid line corresponds to $\xi_e = 0$ and the dashed line represents for 
$\xi_e = 0.15$.

\end{itemize}	


\begin{references}
 
\bibitem{beck} R. Beck, {\em et \, al.}, 
               Ann. Rev. Astron. Astrophys. {\bf 34}, 155 (1996).

\bibitem{kim} K. T. Kim, P. C. Tribble, and P. P. Kronberg, 
             \apj{\bf 379}, (1991); P. P. Kronberg, Rep. Prog. Phys. {\bf 57}, 325 (1994).
 
\bibitem{lanzetta} A. M. Wolfe, K. M. Lanzetta, and A. L. Oren, 
                   \apj{\bf 388}, 17 (1992);
                   P. P. Kronberg, J. J. Perry, and E. L. H. Zukowski, 
                   \apj{\bf 387}, 528 (1992).  

\bibitem{olinto} A. Olinto, in {\em Particle Cosmology},
           edited by K. Sato, T. Yanagida, and T. Shiromizu,
           (Universal Academy Press, Inc., Tokyo, Japan, 1998)

\bibitem{enqvist} K. Enqvist, Int. J. Mod. Phys. D{\bf 7}, 331 (1998);
                  P. L. Biermann, H. Kang, J. Rachen, and D. Ryu,
                  astro-ph/9709252, Les Arcs Proc., p227 (1997)  

\bibitem{barrow} J. D. Barrow, P. G. Ferreira, and J. Silk, \prl{\bf 78}, 3610 (1997);
                 M. Giovannini and M. E. Shaposhnikov, \prl{\bf 80}, 22 (1998);
                 K. Subramanian and J. D. Barrow, \prl{\bf 81}, 3575 (1998).

\bibitem{zeldovich} Y. B. Zeldovich, A. A. Ruzmaikin, and D. D. Sokoloff,
           {\em Magnetic Fields in Astrophysics} (Gordon and Breach, New York, 1983);   
           E. N. Parker, {\em Cosmical Magnetic Fields} (Clarendon Press, Oxford, 1979).

\bibitem{jaffe} W. J. Jaffe, \apj{\bf 241}, 925 (1980);
           A. A. Ruzmaikin, D. D. Sokoloff, and A. Shukurov, MNRAS{\bf 241}, 1 (1989). 

\bibitem{goldman} I. Goldman and Y. Rephaeli, \apj {\bf 380}, 344 (1991);
           D. S. De Young, \apj{\bf 386}, 464 (1992).

\bibitem{goldshmidt} O. Goldshmidt and Y. Rephaeli, \apj {\bf 411}, 518 (1993).

\bibitem{taylor} G. B. Taylor and R. A. Perley, \apj{\bf 416}, 554 (1993); 
                 L. Feretti, $et al.$, Astron. Astrophys. {\bf 302} 680 (1995);
                 Y. Rephaeli and D. E. Gruber, \apj{\bf 333}, 133 (1988);
                 Y. Rephaeli, \apj{\bf 212}, 608 (1977).

\bibitem{kosowsky} A. Kosowsky and A. Loeb, \apj{\bf 469}, 1 (1996).

\bibitem{langer} P. Meszaros, {\em High-energy radiation from magnetized neutron stars},
           (The University of Chicago Press, Chicago, 1992); 
           S. H. Langer, \prd{\bf 23}, 328 (1981);
           I.-S. Suh, $ibid$.{\bf 55}, 4300 (1997);
           L. L. DeRaad Jr., N. D. Hari Dass, and K. A. Milton, \prd{\bf 9}, 1041 (1974);
           $ibid$.{\bf 10}, 1299 (1974).

\bibitem{green} G. Greenstein, Nature, {\bf 223} 938 (1969).
 
\bibitem{oconnell} R. F. O'Connell and J. J. Matese, 
           Nature, {\bf 222}, 649 (1969); Phys. Rev. {\bf 180}, 1289 (1969); 
           \apj {\bf 160}, 451 (1970).

\bibitem{cheng} B. Cheng, D. N. Schramm, and J. W. Truran, \prd{\bf 49}, 5006 (1993);
           B. Cheng, D. N. Schramm, and J. W. Truran, \plb{\bf 316}, 521 (1993);
           B. Cheng, A. V. Olinto, D. N. Schramm, and J. W. Truran,
           \prd{\bf 54}, 4714 (1996).
 
\bibitem{grasso} D. Grasso and H.R. Rubinstein, Astroparticle Phys. {\bf 3} 95 (1995);
           D. Grasso and H. Rubinstein, \plb{\bf 379}, 73 (1996).

\bibitem{kernan} P.J. Kernan, G.D. Starkman, and T. Vachaspati, 
           \prd{\bf 54}, 7207 (1996).

\bibitem{padman} T. Padmanabhan, {\em Structure formation in the Universe},
           (Cambridge University Press, UK, 1993).

\bibitem{kolb} E. W. Kolb and M. S. Turner, {\em The Early Universe}
           (Addison-Wesley, New York, 1990).
           
\bibitem{gusynin} V. P. Gusynin and A. V. Smilga, hep-ph/9807486;
           R. Gepr\"{a}gs, H. Riffert, H. Herold, H. Ruder, and G. Wunner,
           \prd{\bf 49}, 5582 (1994).
 
\bibitem{schwinger} J. Schwinger, {\em Particles, sources and fields},
                    (Addison-Wesley, Redwood City, CA, 1988); 
                    V. Canuto and H.-Y. Chiu, Phys. Rev. {\bf 173}, 1210 (1968).
   
\bibitem{johnson} M. H. Johnson and B. A. Lippmann, Phys. Rev. {\bf 76}, 828 (1949).

\bibitem{landau} L.D. Landau and E.M. Lifshitz, {\em Statistical Mechanics},
               (Clarendon Press, Oxford, 1938).

\bibitem{turner} M. S. Turner and L. M. Widrow, \prd{\bf 37}, 2743 (1988).
 
\bibitem{shapiro} S. L. Shapiro and S. A. Teukolsky, 
                  {\em Black Holes, White Dwarfs, and Neutron Stars}, 
                  (John Wiley $\&$ Sons, New York, 1983).

\bibitem{fassio} L. Fassio-Canuto, Phys. Rev. {\bf 187}, 186 (1969).

\bibitem{kang} H.-S. Kang and G. Steigman, 
               \npb{\bf 372}, 494 (1992).

\bibitem{mathews} R. A. Malaney and G. J. Mathews, 
                  Phys. rep. {\bf 229}, 145 (1993).

\bibitem{sato} K. Kohri, M. Kawasaki, and K. Sato,
               \apj{\bf 490}, 72 (1997).

\bibitem{kimj} J. B. Kim, J. H. Kim, and H. K. Lee,
               Proc. Seventh Asian-Pacific Regional Meeting of the IAU, 1996
               (astro-ph/9701011).

\bibitem{copi} C. J. Copi, D. N. Schramm, and M. S. Turner,
               \prl{\bf 75}, 3981 (1995).

\bibitem{isuh} In-Saeng Suh and G. J. Mathews, in preparation.
\end{references}
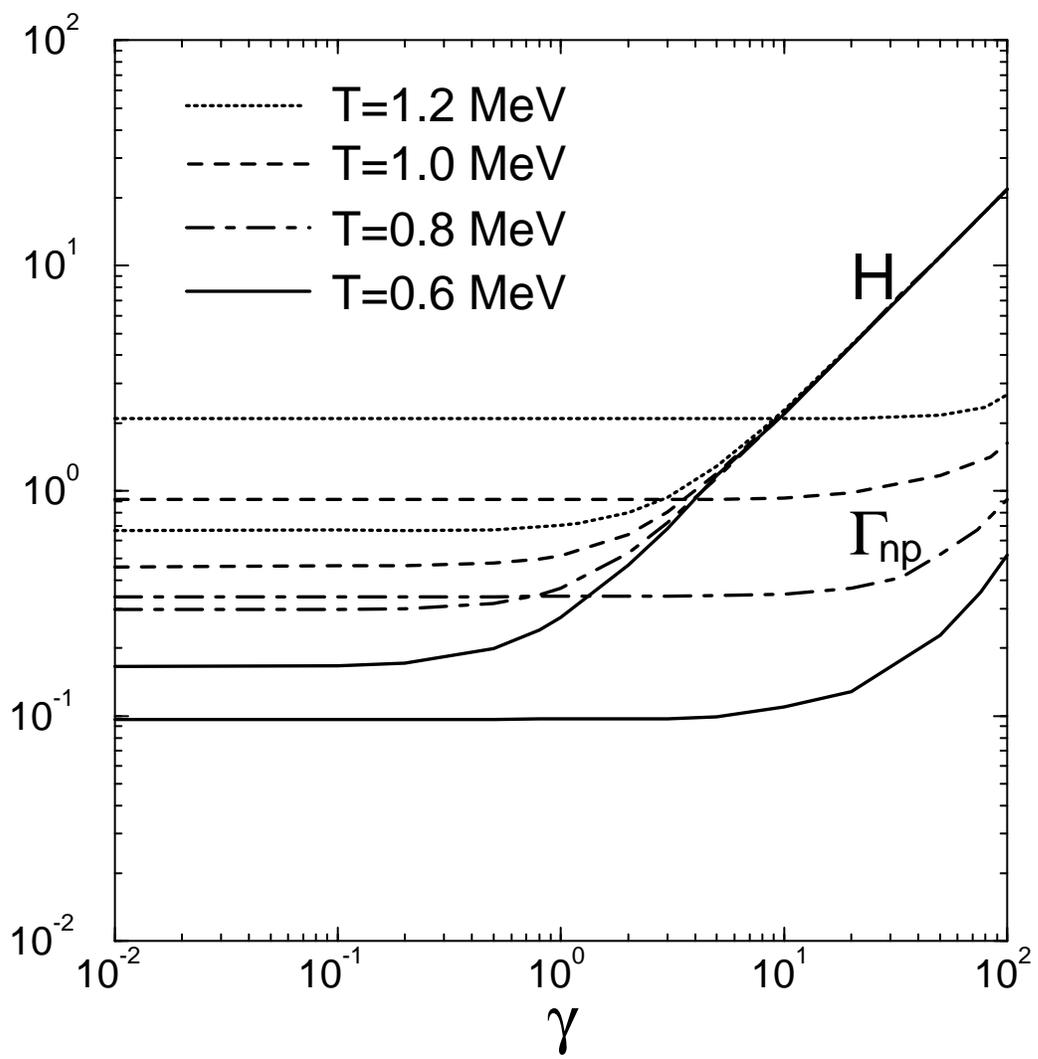
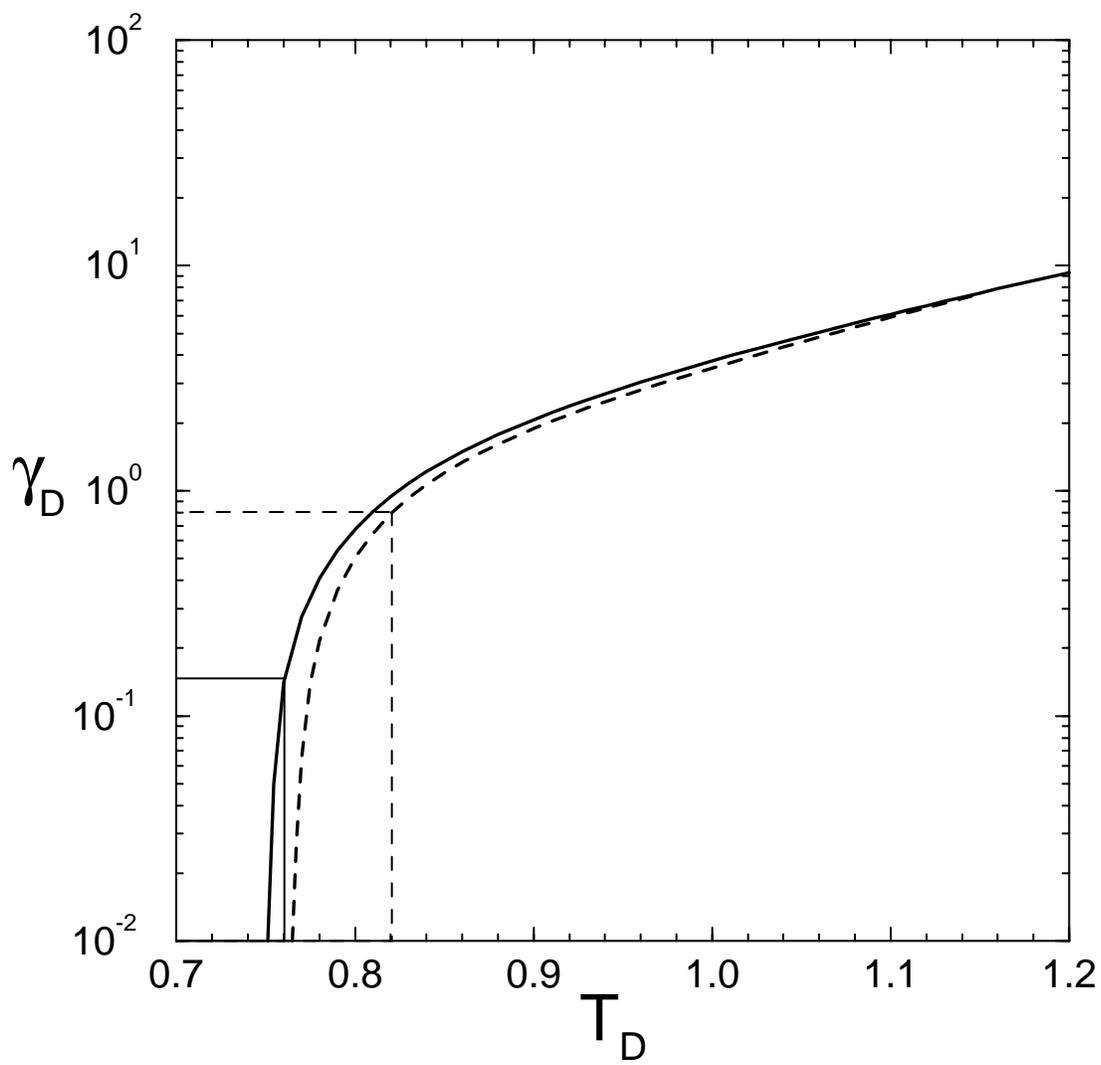
\end{document}